 \documentclass[page-classic]{epl2} 

\newcommand{\marge}[1]{\marginpar{}}  
\newcommand{\Sl}[1]{{}}           
\newcommand{\beq}[1]{\Sl{#1}\begin{equation}\if#1\empty\else\label{#1}\fi}
\newcommand{\eeq}{\end{equation}}
\newcommand{\beqa}[1]{\Sl{#1}\begin{eqnarray}\if#1\empty\else\label{#1}\fi}
\newcommand{\eeqa}{\end{eqnarray}}

\newcommand{\nm}{\nonumber\\}
\newcommand{\Eq}[1]{Eq.(\ref{#1})}

\newcommand{\la}{\langle}
\newcommand{\ra}{\rangle}

\newcommand{\E}{\mathsf E}

\title{Nonlinear diffusion from Einstein's master equation}
\shorttitle{Nonlinear diffusion} 

\author{J.P. Boon \thanks{E-mail: \email{jpboon@ulb.ac.be}}
  \and J.F. Lutsko \thanks{E-mail: \email{jlutsko@ulb.ac.be}} }
\shortauthor{J.P. Boon and J.F. Lutsko}

\institute{                   
 Physics Department CP 231, Universit\'e Libre de Bruxelles, 1050 - Bruxelles, Belgium
}

\abstract{
We generalize Einstein's master equation for random walk processes
by considering that  the probability for a particle at position $r$ to
make a jump of length $j$ lattice sites, $P_j(r)$  is a 
functional of the particle  distribution function $f(r,t)$.
By multiscale expansion, we obtain a generalized advection-diffusion equation.
We show that the power law $P_j(r) \propto f(r)^{\alpha - 1}$ (with $\alpha > 1$)
follows from the requirement that the generalized equation admits of scaling solutions ($ f\left( r;t\right) = t^{-\gamma }\phi \left( r/t^{\gamma }\right) $).
The solutions have a $q$-exponential form and are found to be in agreement with the results of Monte-Carlo simulations, so providing a microscopic basis validating the nonlinear diffusion equation.
Although its hydrodynamic limit is equivalent to the phenomenological 
porous media equation, there are extra terms which, in general, cannot be
neglected as evidenced by the Monte-Carlo computations.}

\begin{document}

\maketitle

\section{Introduction}
A standard procedure to describe the microscopic mechanism of a diffusion process, is to consider a test particle executing a random walk on some substrate. The idea goes back  to Einstein who, in one of his celebrated 
1905 articles \cite{einstein}, showed how the diffusion equation follows
from a mean-field description written  in terms of the probabilities that 
the particle performs elementary displacements at each time step.
The distribution function $f(r,t)$, that is the probability that, given the particle was initially at $r=0$ at $t=0$, it will be at position $r$ at time $t$ (for $t$ large compared to the duration of an elementary displacement) is obtained as the 
solution to the Fokker-Planck equation for diffusion, and one finds
that, in the the long-time limit, $f(r,t)$ is Gaussianly distributed in 
space \cite{feller}. This result had been known since Fick's law was
established for diffusion; what was new in Einstein's work was the 
microscopic content, in particular the expression of the diffusion coefficient in terms of the particle velocity autocorrelation function, a form that was further
generalized to the general class of transport coefficients known since the
1960's as the Green-Kubo coefficients \cite{green-kubo}. 

The classical diffusion equation has been extensively used and successfully
applied to a large class of phenomena (ranging from particle dispersion
in suspensions to diffusion of innovations in social networks) and is indeed
applicable as long as the system responds linearly to a change in the 
quantity that is being transported. But the linear response hypothesis
doesn't hold in more complicated situations such as when there is an interactive 
process between the particle and the substrate, or in heterogeneous media.
In the 1930's, a nonlinear diffusion equation was proposed on a purely phenomenological basis devised in particular to describe diffusive transport
in porous media, hence the name {\em porous media equation} \cite{muskat}
\beq{PM_Eq}
\frac{\partial}{\partial t} f(r;t)\,=\,  
 D \,\frac{\partial^2}{\partial {r^2}}f^{\alpha}(r;t)\,,
\eeq
where $D$ is the diffusion coefficient. This equation, when
generalized with an advective term,  has a $q$-Gaussian solution \cite{PlastinoPlastino}
and exhibits the interesting feature that the scaling $\la r^2 \ra \propto t^{\gamma}$ 
can be non-classical ($\gamma \neq 1$). \footnote{When $\alpha = 1$, \Eq{PM_Eq} is the usual diffusion equation with a Gaussian solution \cite{feller} and classical scaling ($\gamma = 1$).}
It was not until the 1990's that a more fundamental basis was proposed
for the (generalized) porous media equation using various statistical mechanical approaches. The reason for the  various approaches can be found in the variety of problems where non-classical 
(non-Gaussian) distributions are observed: transport in porous media, viscous fingering, information diffusion in social networks or in the internet, financial market 
distributions, ... . The proposed approaches use formulations such as the generalized 
entropy \cite{TsallisBukman}, the Langevin equation \cite{Borland,AnteneodoTsallis},
the master equation \cite{CuradoNobre}, the nonlinear response \cite{LutskoBoon},
the escort distribution \cite{AbeThurner}, or the generalized generating function 
\cite{BoonLutsko}.   

\section{Generalized master equation}
Here we use the microscopic approach by going back to Einstein's original derivation based on the random walk. 
For simplicity consider a one-dimensional lattice where
the particle hops to the nearest neighboring site (left or right) in one time step, 
a process described by the discrete equation
\beq{RW_Eq}
n(r;t+1) = \xi_{-} n(r+1;t) + \xi_{+} n(r-1;t)\;,  
\eeq
where the Boolean variable $n(r;t) = \{0,1\}$ denotes the occupation at time $t$ of the site located at position $r$ and $\xi_{\pm}$ is a Boolean random variable controlling the particle jump between neighboring sites ($\xi^+ + \xi^- = 1$). The mean field description
follows by ensemble averaging \Eq{RW_Eq}. With $\la n(r;t) \ra = f(r;t)$ and
$\la \xi_j \ra = P_j$ (using statistical independence of $\xi$ and $n$), and extending the possible jump steps over the whole lattice,  one obtains Einstein's master equation \cite{einstein}
\beq{Ein_Eq}
 f(r;t + \delta t)\,=\,\sum_{j=-\infty}^{+\infty} P_j(r - j\delta r ; t) \, f(r - j\delta r ; t) \;,
 \eeq
where $P_j(\ell)$ denotes the probability that the walker at site $\ell$ make a jump of $j$ sites  \footnote{In Einstein's formulation the particle jumps are restricted to symmetrical displacements, i.e.  $P_{+j}=P_{-j}$.}. Using the normalization:
$1\,=\,P_{0}\left( \ell;t\right)  + \sum_{j\neq 0}P_{j}\left( \ell;t\right) $\,,
\Eq{Ein_Eq} takes the form of a Boltzmann like difference equation
\beqa{MF_Eq}
f\left( r;t+\delta t\right) -f\left( r;t\right) = 
\sum_{j = 0} [P_{j}\left( r - j {\delta r};t\right) f\left( r-j\delta r;t\right) - 
P_{j}\left( r;t\right) f\left( r;t\right)] \,,
\eeqa
which simply describes the rate of change of the particle distribution as the 
difference between the incoming and outgoing fluxes at location $r$.

As discussed in the introductory section, in more complex situations, when the linear
response hypothesis is no longer valid, one observes non-Gaussian behavior, i.e. the long-time dynamics is different from that described by the classical Fokker-Planck equation (or the usual advection-diffusion equation). At the level of the mean field description, the breakdown of linear response means that the particle motion depends  on the occupation probability in a non-trivial way. The jump probability then becomes a functional of the particle distribution function, and Einstein's equation describing the space-time evolution of the particle motion must be generalized in order to account for the functional dependence. So introducing  
$P_{j}\left( \ell ;t\right) = p_{j}\,F\left( f\left( \ell \delta r;t\right) \right)\,,$ 
with $j \neq 0$ and where  $p_j $ is a given distribution of displacements (e.g.
$p_j \propto \exp -j$) in (\ref{MF_Eq}), we obtain the generalized master equation
\beqa{GM_Eq}
 f\left( r;t+\delta t\right) -f\left( r;t\right) \,=\,  
 \sum_{j = 0} p_j[F\left(f( r - j {\delta r};t)\right) f\left( r-j\delta r;t\right)\,-\,  
  F\left(f (r;t)\right) f\left( r;t\right) ]\,.
\eeqa

\section{Generalized diffusion equation}
Along the same lines as the classical diffusion equation is obtained from
Einstein's master equation, we perform a multi-scale expansion of the
generalized master equation (\ref{GM_Eq}) using 
an expansion of the time and space derivatives of the form%
\begin{eqnarray}
\label{expan_rt}
\frac{\partial }{\partial t} &=&\epsilon \frac{\partial^{\left( 1\right)}}{\partial t}
+\epsilon ^{2}\frac{\partial^{\left( 2\right) }}{\partial t}+... \,, \nm 
\frac{\partial }{\partial r} &=&\epsilon \frac{\partial^{\left( 1\right) }}{\partial r}
+\epsilon ^2 \frac{\partial^{\left( 2\right) }}{\partial r} +... \,, 
\end{eqnarray}%
and a corresponding expansion of the distribution as%
\begin{equation}
\label{expan_f}
f\left( r;t\right) =f_{0}\left( r;t\right) +\epsilon f_{1}\left( r;t\right) +... \,.
\end{equation}%
To first order, we have
\begin{eqnarray}
\label{1st_eq}
{\cal O}(\epsilon^{1})\,:\qquad
\frac{\partial^{\left( 1\right)}}{\partial t} f_{0}\left( r;t\right) =-\left( J_{1}\frac{%
\delta r}{\delta t}\right)  \frac{\partial^{\left( 1\right) }}{\partial r} F\left(
f_{0}\left( r;t\right) \right) f_{0}\left( r;t\right) \,,
\end{eqnarray}%
and to second order%
\begin{eqnarray}
{\cal O}(\epsilon^{2})\,: \qquad
\frac{\partial^{\left( 1\right) }}{\partial t} f_{1}\left( r;t\right) +
\frac{\partial^{\left( 2\right)}}{\partial t} f_{0}\left( r;t\right) +\frac{1}{2}\left( \delta
t\right) \frac{\partial^{\left( 1\right) 2}}{\partial t^2} f_{0}\left( r;t\right) = \nm
-\left(J_{1}\frac{\delta r}{\delta t}\right)\frac{\partial^{\left( 1\right) }}{\partial r}\left(
\left. \frac{dgF\left( g\right) }{dg}\right\vert _{g=f\left( r;t\right)
}f_{1}\left( r;t\right) \right)
-\left( \frac{\delta r}{\delta t}J_{1}\right) \frac{\partial^{\left( 2\right) }}{\partial r}
F\left( f_{0}\left( r;t\right) \right) f_{0}\left( r;t\right) 
\nonumber \\
+\frac{1}{2}\left( \frac{\left( \delta r\right) ^{2}}{\delta t}%
J_{2}\right)\frac{\partial^{\left( 1\right)2 }}{\partial r^2}F\left( f_{0}\left( r;t\right)
\right) f_{0}\left( r;t\right) \,,
\end{eqnarray}%
where $J_{n}$ denotes the moments  $J_{n}=\sum_{j\neq 0}j^{n}p_{j}$.
Resummation of  these results (see \cite{full_paper} for details)
yields the hydrodynamic limit of the generalized master equation
\beqa{GD_Eq}
 \frac{\partial }{\partial t}f\left( r;t\right) +C\frac{\partial }{\partial r}%
\left[F\left( f\left( r;t\right) \right) f\left( r;t\right)\right]  =  \nm
D\frac{\partial^2}{\partial{r}^{2}}
\left[F\left( f\left( r;t\right) \right) f\left( r;t\right)\right]
+\,\frac{1}{2}\left( C^{2}\delta t\right) \frac{\partial }{\partial r} {\E}(r;t) \,.
\eeqa%
This result is the generalized diffusion equation
where $C$ and $D$ are the drift velocity and the diffusion coefficient respectively
\begin{eqnarray}
C\,=\,\frac{\delta r}{\delta t} \sum_{j\neq 0} j p_{j} \,, \qquad
D \,=\, \frac{\left( \delta r\right)^{2}}{2\delta t}
\left(\sum_{j\neq 0}j^{2}p_{j}  -  \left( \sum_{j\neq 0}j p_{j}\right)^{2}\right) \,,
\end{eqnarray}%
and
\begin{eqnarray}
{\E}(r;t) \,=\,\frac{\partial }{\partial r}\left[F(f\left( r;t\right)) f\left( r;t\right)\right] \,-\, 
\left(\left\vert\frac{dgF\left( g\right) }{dg}\right\vert _{g=f\left( r;t\right) }\frac{\partial }{\partial r}\left[F(f\left( r;t\right)) f\left( r;t\right) \right]\right) \;.
\end{eqnarray}%
When there is an external force acting on the particle, \Eq{GD_Eq} is further
generalized as discussed elsewhere \cite{full_paper}.

\section{Scaling solution}
Under which conditions is there a scaling solution to the generalized equation?
For simplicity consider the diffusion equation with no drift, i.e. the jump probability
is space symmetrical and consequently the first moment $J_1 = 0$ and $C = 0$,
and (\ref{GD_Eq}) reduces to
\begin{equation}
\label{SD_Eq}
\frac{\partial }{\partial t}f\left( r;t\right) =
D\frac{\partial^{2}}{\partial  {r}^{2}}[F\left(f\left( r;t\right) \right) f\left( r;t\right)] \,.
\end{equation}%
Assuming that 
$f\left( r;t\right) =t^{-\gamma }\phi \left( r/t^{\gamma }\right) =t^{-\gamma
}\phi \left( x\right) $, and expressing the time and space derivatives in terms of $x$,
\Eq{SD_Eq} is rewritten as (see \cite{full_paper} for details)
\begin{equation}
\label{x_D_Eq}
-\gamma \frac{d}{dx}x\phi \left( x\right) =Dt^{1-2\gamma }\frac{d^{2}}{dx^{2}%
}F\left( t^{-\gamma }\phi \left( x\right) \right) \phi \left( x\right) \,.
\end{equation}%
The time-dependence on the right can only be eliminated if $F(g)=g^{\eta } = t^{-\eta \gamma }\,\phi^{\eta}$ for some number 
$\eta$, and hence $1=t^{1-2\gamma }t^{-\eta \gamma }$, i.e.
\begin{equation}
\label{gamma}
\gamma =\frac{1}{2+\eta }\,.
\end{equation}%
Thus, when $\eta \neq 0$, this describes anomolous diffusion: 
$\left\langle r^2\right\rangle \sim t^{\frac{2}{2+\eta }}$. 
In this case, \Eq{x_D_Eq} becomes
\begin{equation}
D\frac{d^{2}}{dx^{2}}\phi ^{1+\eta }\left( x\right) +\frac{1}{2+\eta }\frac{d%
}{dx}x\phi \left( x\right) =0 \,,
\end{equation}%
and admits a q-exponential solution (see \cite{full_paper} for details)
\beqa{hom_sol}
\phi (x) = \left( \frac{1+2\eta }{1+\eta }B\right) ^{\frac{1}{\eta }}\left[ 1-\frac{%
\eta }{2\left( 2+\eta \right) \left( 1+2\eta \right) BD}x^2 \right] ^{\frac{1}{%
\eta }} \,,
\eeqa%
where $B$ is an integration constant. With $\eta = 1-q$, and returning to the original space and time variables, (\ref{hom_sol}) takes the canonical $q$-exponential form
\beq{q_sol}
 f(r;t)\,=\,B_q t^{-\frac{1}{3-q}}\left[ 1- (1-q) M_q \,\frac{r^2}{D\,t^{\frac{2}{3-q}}}%
                 \right]^{\frac{1}{1-q}}
\eeq
with
\beqa{MqBq}     
 B_q  = \left[\left(1 + \frac{1-q}{2-q}\right) B \right]^{\frac{1}{1-q}} \,, \qquad
 M_q^{-1} = 2(3-q)\left(3-2q \right) B \,.
 \eeqa%
So, with no drift and no external field, the generalized random walk model describes anomalous diffusion with q-distributions \footnote{One verifies straightforwardly that
for $q \rightarrow 1$, one retrieves the classical Gaussian distribution.}.
It also follows from (\ref{q_sol}) that the distribution of the values taken by $f(r;t)$ 
at any fixed value of time has the form of a power law \cite{GrosfilsBoon};
\beqa{f_distrib}
 {\cal P}({\tilde f}) = \, \int^{\infty}_{-\infty} dr \,
 D^{-\frac{1}{2}}\,t^{-\frac{1}{3-q}}\, \delta({\tilde f}(r;t) - {\tilde f}) 
 \sim  \frac {{\tilde f}^{-q}} {\sqrt {1-{\tilde f}^{1-q}}}\,.
\eeqa

An important result of the present analysis is that the power law dependence of
the transition probability, $P_j = p_j\,F(f)$ with $F(f) = f^{\eta}$, is not introduced as
an ansatz, but follows from the demand for a scaling (or self-similar) solution to
the generalized diffusion equation.

Now introducing the power law dependence $F(f) =  f^{\eta}$ (with $\eta \geq 0$
for normalization $\sum_{j} P_j = 1$)  in the generalized equation (\ref{GD_Eq}), we obtain (with $\eta = \alpha -1$)
\beqa{GD_alpha}
 \frac{\partial }{\partial t}f\left( r;t\right) +C\frac{\partial }{\partial r}%
 f^{\alpha}\left( r;t\right)  = 
D\frac{\partial^2}{\partial{r}^{2}} f^{\alpha}\left( r;t\right)
+\,\frac{1}{2}\left( C^{2}\delta t\right) \frac{\partial }{\partial r} {\E}(r;t) \,,
\eeqa%
with
\beqa{extra_term}
{\E}(r;t) \,=\,\left(1-\alpha f^{\alpha -1}\left( r;t\right)\right)
\frac{\partial }{\partial r} f^{\alpha}\left( r;t\right)  \,.
\eeqa%
Comparison of \Eq{GD_alpha} and \Eq{PM_Eq} shows that
the two equations are the same in the absence of drift ($C = 0$). Even with
nonzero drift, this corresponds to a generalized porous media
equation when the second term on the r.h.s. of
(\ref{GD_alpha}) vanishes, i.e. for $\delta t \rightarrow 0$.
So the phenomenological generalized porous media equation is an approximation which can be obtained in the hydrodynamic limit from the generalized master equation with a power law dependence for the transition probability
(and in the absence of external force \cite{full_paper}). However \Eq{GD_alpha} contains an additional term which, in general, cannot be neglected (see next  section).

\begin{figure}
\onefigure[scale=0.4]{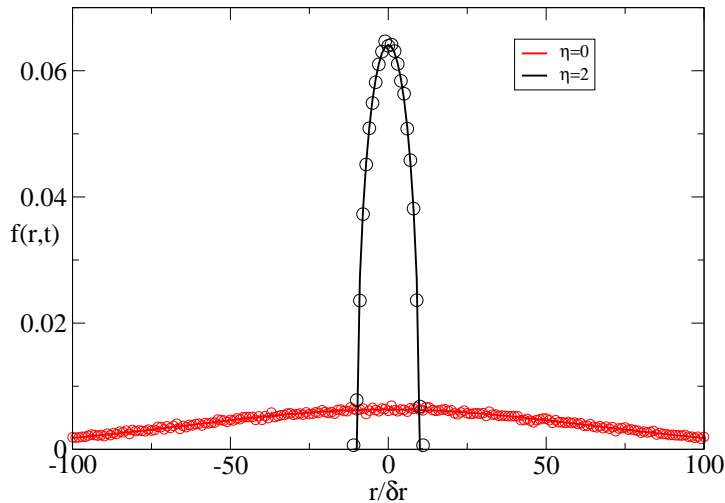}
\caption{(Color online) Generalized diffusion with no drift: the distribution function
$f(r ; t=2000$ time steps) obtained from Monte-Carlo simulations (symbols) and the 
solution of the generalized diffusion equation (\ref{q_sol}) (solid lines) for 
$\alpha = 3$ (upper curve) and $\alpha = 1$ ($10^5$ walkers; initial condition: 
$f(r,t=0) = \delta (r)$).}
\label{fig_no_drift}
\end{figure}

\section{Microscopic simulations}
Monte-Carlo simulations are performed with the  generalized master equation
(\ref{GM_Eq}) using the power law dependent jump probabilities and prescribed
$p_j$ distributions: $p_j = \frac{1}{5}$ for $j = [-2,+2 ]$  (space
symmetrical jumps and, so, $C$=0) and
$p_j = \frac{j+3}{15}$ for $j = [-2,+2]$ (space asymmetrical jumps and
so with non-zero drift velocity), and the results are compared
with the numerical solution of the generalized diffusion equation
(\ref{GD_alpha}). Figure \ref{fig_no_drift}  illustrates the case without drift for $\eta = 2$ ($\alpha = 3$ and $q = -1$) showing perfect agreement between the Monte-Carlo data and the $q$-exponential solution (\ref{q_sol}); 
for comparison the classical Gaussian result  ($\eta = 0$, $q=1$) is also shown. 

\begin{figure}
\onefigure[scale=0.4]{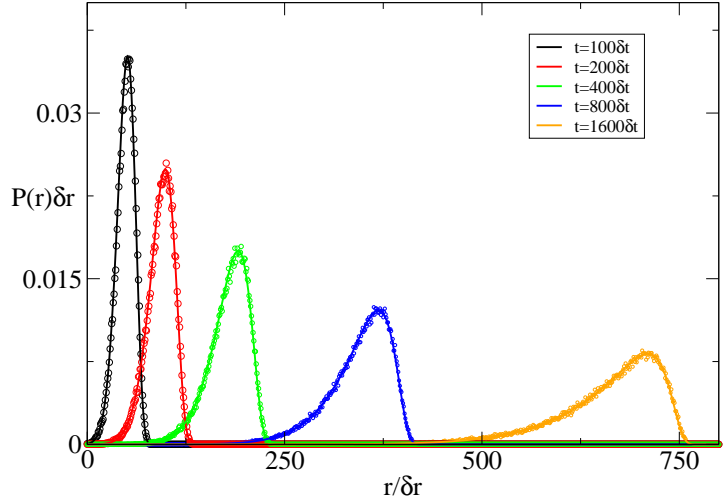}
\caption{(Color online) Generalized diffusion with drift: the distribution function 
$f(r ; t = 100$ to $1600$ time steps) obtained from Monte-Carlo simulations 
(symbols) and the numerical solution of the generalized diffusion equation 
(\ref{GD_alpha}) (solid lines) for $\alpha = 1.1$, i.e. $q=0.9$  ($10^5$ walkers; 
initial condition:  $f(r,t=0) = \delta (r)$).}
\label{fig_drift_01}
\end{figure}

\begin{figure}
\onefigure[scale=0.4]{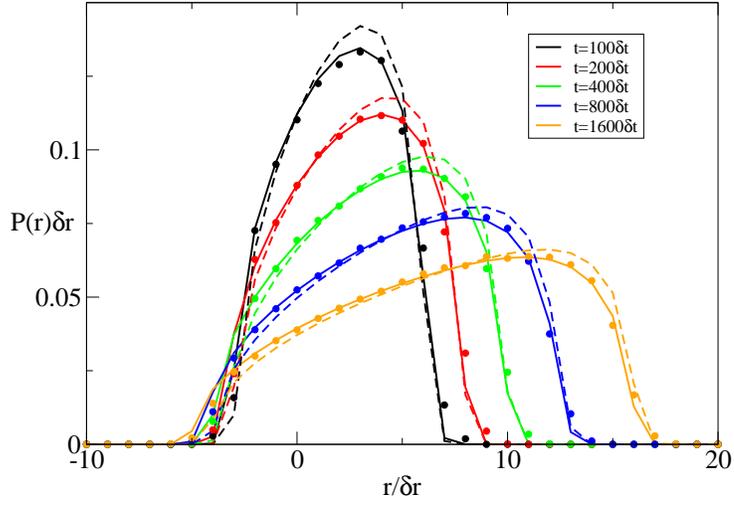}
\caption{(Color online) Same as Fig.\ref{fig_drift_01} for $\alpha = 2$, i.e. $q=-1$. 
Dashed lines: see text}
\label{fig_drift_20}
\end{figure}

Two examples with drift are given in Figs.\ref{fig_drift_01} and \ref{fig_drift_20} for 
$\alpha = 1.1$ ($q=0.9$) and $\alpha = 2$ ($q=-1$) respectively showing excellent
agreement between the simulation data and the solution of the nonlinear equation.
We also computed the solution of the generalized diffusion equation
without the extra term  $E(r;t)$  for the value $\alpha = 2$; the results are given by the dashed lines in Fig.\ref{fig_drift_20}.
The systematic discrepancy with the simulation  results gives clear evidence that
the term given by (\ref{extra_term}) in the  generalized equation (\ref{GD_alpha}) 
cannot be neglected. To the best of our knowledge the present results provide the
first microscopically based demonstration of the nonlinear diffusion equation. 
Further results, including the case where the transition probabilities have
full spatial dependence (i.e. not only on the distribution at the originating location) and  the generalization with an external force (i.e. the nonlinear
advection-diffusion equation) are discussed in \cite{full_paper}.

\acknowledgments
{\bf Acknowledgments.} 
The work of JFL was supported in part by the European Space Agency
under contract number ESA AO-2004-070.

\end{document}